# Price change prediction of ultra high frequency financial data based on temporal convolutional network

Wei Dai[a,b,c], Yuan An[a,b,c], Wen Long[a,b,c,*]

[a]*School of Economics and Management, University of Chinese Academy of Sciences, Beijing, 100190, P.R.China*
[b]*Research Center on Fictitious Economy and Data Science, Chinese Academy of Sciences, Beijing, 100190, P.R.China*
[c]*Key Laboratory of Big Data Mining and Knowledge Management, Chinese Academy of Sciences, Beijing, 100190, P.R.China*

E-mail address: bishengky@163.com; longwen@ucas.ac.cn.

**Abstract**

Through in-depth analysis of ultra high frequency (UHF) stock price change data, more reasonable discrete dynamic distribution models are constructed in this paper. Firstly, we classify the price changes into several categories. Then, temporal convolutional network (TCN) is utilized to predict the conditional probability for each category. Furthermore, attention mechanism is added into the TCN architecture to model the time-varying distribution for stock price change data. Empirical research on constituent stocks of Chinese Shenzhen Stock Exchange 100 Index (SZSE 100) found that the TCN framework model and the TCN (attention) framework have a better overall performance than GARCH family models and the long short-term memory (LSTM) framework model for the description of the dynamic process of the UHF stock price change sequence. In addition, the scale of the dataset reached nearly 10 million, to the best of our knowledge, there has been no previous attempt to apply TCN to such a large-scale UHF transaction price dataset in Chinese stock market.

*Keywords:* Temporal convolutional network; Ultra high frequency; Financial prediction

## 1. Introduction

With the advent of the big data era, as an important artificial intelligence technology, machine learning methods have received more and more attention from researchers, and various deep learning methods have been increasingly used in financial market analysis. In [1], Chanakya Serjam and Akito Sakurai chose SVR model to predict price movements within 1 minute in the currency market, and obtained excellent performance in simulated trading. Ref [2] compared the performance of linear discriminant analysis, Logit, artificial neural network, random forest and SVM in predicting the direction of daily stock index movement. Ref [3] systematically analyzed the potential of deep neural networks in predicting the stock market movement at high frequencies and found that the DNN method can extract information from the residuals of the autoregressive model. Ref [4] used LSTM network for financial market forecasting in order to more effectively identify the temporal information of sequential data.

Ref [5] proposed a machine learning framework to describe the dynamics of high-frequency limit order books in financial stock market and automatically predict indicators such as intermediate prices and spreads in real time. By using attribute vectors of different levels (such as price and transaction volume) to characterize data in the limit order book, the proposed framework builds a learning model for each indicator based on multiple types of support vector machines. Empirical research shows that the feature extraction performed by the proposed framework is effective for short-term price trend prediction. The empirical research on real data sets also shows that the feature set defined in advance is reasonable and effective. In addition, a large number of comparison experiments indicate that, compared with the three multi-kernel SVMs in [6], the multi-class SVM modeling technology in [5] has excellent performance in terms of accuracy and recall while reducing training costs.

Ref [7] proposed a novel time-aware neural feature bag model (bag-of-features, BoF), which is suitable for time series forecasting using high-frequency limit order data. Two independent radial basis functions and cumulative layer sets are used in the time BoF to characterize the short-term behavior and long-term dynamic characteristics of the time series. Any other neural layer (such as a feature conversion layer) or classifier (such as a multi-layer perceptron) can be used in combination with the BoF method. Ref [8] used text disclosure information to construct a deep learning model for predicting corporate bankruptcy, and conducted an empirical study on a corporate bankruptcy data set containing 11,827 listed companies in the United States, and the constructed model achieved outstanding prediction performance. By comparing with different deep learning frameworks, the study found that the simple average embedding calculation is more effective than convolutional neural network. Ref[9] used logistic regression, support vector machines, and random forest models to evaluate the importance of 44 technical indicators calculated on 88 stocks. According to a pre-set threshold, the lowest ranked technical indicators were excluded. The remaining indicators are divided into different groups. Based on the previously calculated results, the research selected the most important technical factors from each technical indicator group to form the input of the deep generation model. The deep generative model is composed of a market signal extractor and an attention mechanism layer. The experiment results prove the effectiveness of this method.

Ref [10] combined two deep learning structures by adding a long short-term memory network (LSTM) to a convolutional neural network (such as a pre-trained VGG16), which results in a more accurate prediction for the fluctuation of gold. Ref [11] proposed a simple and effective neural layer for the data normalization task, which can adaptively normalize the input time series in consideration of the data distribution. Different from the traditional normalization method, Ref [11] constructed a neural network model to learn how to perform normalization for a given task, instead of using a fixed normalization scheme. Experiments performed on large-scale order book data verify the effectiveness of the method. Ref [12] combined the return prediction in the process of portfolio construction with two machine learning models (namely random forest (RF) and support vector regression (SVR)) and three deep learning models (namely LSTM neural network, deep multi-layer perceptron (DMLP) and convolutional neural network). Specifically speaking, the research first applies these forecasting models to stock pre-selection before investment portfolio formation, and then incorporates the forecast results into portfolio optimization model.

Recently, temporal convolutional networks(TCN)[13, 14] also have exhibited excellent performance in financial applications. This study analyzes the relevant literature searched from Google Scholar. Specifically, this study separates search terms into two strings. The first string is "temporal convolutional network" and the second string is one of "ultra high frequency", "trade by trade", "transaction by transaction" and "tick data". Hence, four combinations of strings are searched in Google Scholar. To our surprise, there has been no previous study that tried to apply TCN architecture to large-scale ultra high frequency(UHF) transaction price data. This study concentrates on UHF transaction price change data and takes the characteristic of discrete price change into consideration to find a more suitable model to describe the dynamic distribution for the UHF data via utilizing TCN architecture.

## 2. Methodology

*2.1. UHF data characteristics*

As the stock price changes (see Eq. (1)) can only be several times of the minimum pricing unit, it is more suitable to describe the distribution of UHF price changes by a discrete function, which is also one of the main differences between UHF financial data and non-UHF financial data.

$$\Delta P_i = P_i - P_{i-1} \tag{1}$$

$$\begin{aligned}
&Class\ 0: \Delta P_i \leq -0.02 \\
&Class\ 1: \Delta P_i = -0.01 \\
&Class\ 2: \quad \Delta P_i = 0 \\
&Class\ 3: \Delta P_i = 0.01 \\
&Class\ 4: \Delta P_i \geq 0.02
\end{aligned} \tag{2}$$

$$\pi_i = \begin{cases} \pi_{i0}, & \Delta P_i \leq -0.02, \\ \pi_{i1}, & \Delta P_i = -0.01, \\ \pi_{i2}, & \Delta P_i = 0, \\ \pi_{i3}, & \Delta P_i = 0.01, \\ \pi_{i4}, & \Delta P_i \geq 0.02, \end{cases} \tag{3}$$

Figure 1 and Figure 2 are histograms of price change distribution for Ping An Bank Co., Ltd. (000001.SZ) and CHINA VANKE CO., LTD. (000002.SZ). It can be seen very intuitively that in the transaction data of these two stocks, transactions with price change 0 yuan account for more than 70 percent, and price change of +0.01 yuan or -0.01 yuan also account for more than 10 percent respectively. The difference between the two distribution diagrams is that the ratio of the price change of +0.02 yuan or -0.02 yuan in Figure 2 is higher than that in Figure 1. Since the distribution of these two stocks is considerable representative among the constituent stocks of Shenzhen Stock Exchange 100 Index, this study next divides the price changes of UHF trading into five categories (see Eq.(2)), and uses deep learning models to predict the discrete conditional distribution of each transaction (see Eq.(3)).

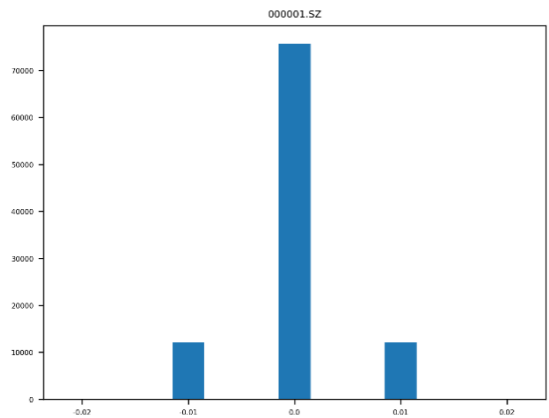

Fig. 1. Distribution of price changes for stock 000001.SZ

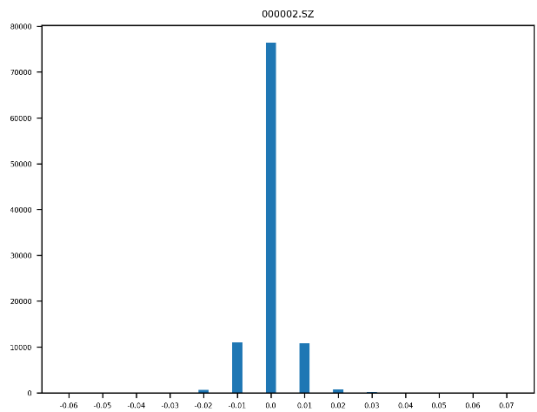

Fig. 2. Distribution of price changes for stock 000002.SZ

### 2.2. TCN network structure

Nowadays, recurrent neural network (RNN) is one of the most commonly used deep learning network structure for modeling sequential data. However, in some recent studies, temporal convolutional networks (TCN)[13, 14] have demonstrated its ability of storing and memorizing information, and in many sequential modeling tasks, its performance has outperformed the benchmark RNN models, which provides financial researchers with more choices when dealing with financial time series data.

### 2.3. Dilated Convolution

First of all, TCN adopts a causal convolution structure to ensure that future time information will not be used in the prediction process at the current time, that is, the convolution output corresponding to time $T$ can only use information at time $T$ or before time $T$. In addition, the TCN structure often increases the perceptual field of the convolutional neural network through dilated convolution[13], so that it can use the historical information of a longer time range. A *d* parameter is usually used to represent expansion rate in the corresponding dilated convolution layer. When *d = 1*, the dilated convolution is equivalent to the general causal convolution.

*2.4. Residual module*

The memory capacity of TCN model depends on the depth of its network layers. However, the prediction performance of a deep neural network may decline with the increase of network layers. The residual network ResNet (Residual Network) proposed by He Kaiming et al.[15] makes deep network structure easier to optimize by introducing residual mapping, which greatly facilitates the application of deep network structures. Therefore, ordinary convolution calculations are often replaced by residual network modules in TCN framework.

## 3. Experiment

*3.1. Experiment data*

This study conducts empirical research on nearly all the constituent stocks of the Shenzhen Stock Exchange 100 Index released on December 31, 2016. The Shenzhen Stock Exchange 100 Index is a product index compiled by taking into account factors such as market capitalization, liquidity, and historical returns, and selecting 100 stocks with better overall performance from all stocks traded in the Shenzhen Stock Exchange. The SZSE 100 Index needs to satisfy that the number of SMEs in the constituent stocks is not less than 10, so the constituent stocks of the SZSE 100 Index cover multiple levels of the stock market. This paper selects the first 100,000 UHF trading data of each stock during the continuous auction period in 2017, and divides it into a training set and a test set according to a ratio of 7:3. In addition, since TIANJIN ZHONGHUAN SEMICONDUCTOR CO., LTD. (002129.SZ) was suspended for all the year of 2017, this study removed it from the sample, so the experiment dataset contains 9,900,000 UHF trading transactions from 99 SZSE 100 constituent stocks.

*3.2. Experiment result*

In addition to TCN model, this study also utilizes LSTM model and adds attention mechanism layer into the TCN to construct a TCN(attention) structure to predict the discrete distribution of UHF price changes. Furthermore, the nine GARCH family models have also been applied to the prediction of the logarithmic return distribution of the 99 SZSE 100 constituent stocks. For the UHF price change models based on LSTM, TCN and TCN(attention) structure, the loss function is set as the cross-entropy loss(see Eq.(4)). In the experiment on 99 stocks, fixed hyperparameters are chosen for LSTM, TCN and TCN (attention) model.

$$L = -\frac{1}{n}\sum_{i=1}^{n}\sum_{k=0}^{4} y_{ik} \cdot ln\pi_{ik} \qquad (4)$$

After the GARCH family models being applied to the log return data of UHF transactions, the distribution of the log return of the *i*-th transaction can be obtained. And then, the distribution of UHF stock price changes can be calculated. However, the probability of price change less than -0.02 or lager than 0.02 predicted by the GARCH family models are too small. In order to facilitate the comparison between the six GARCH family models and the three models based on deep learning framework, the five-class classification results are transformed into three-class classification results according to the following equation.

$$\begin{aligned} &Class\ A: &\Delta P_i \leq -0.01 \\ &Class\ B: &-0.01 < \Delta P_i < 0.01 \\ &Class\ C: &\Delta P_i \leq 0.01 \end{aligned} \qquad (5)$$

The GARCH family models selected in this study include SGARCH-norm, SGARCH-std, SGARCH-sstd, EGARCH-norm, EGARCH-std, EGARCH-sstd, GJR-GARCH-norm, GJR-GARCH -std, GJR-GARCH-sstd[16], amount to 9 models. Since the three models of GJR-GARCH specification have numerical calculation problems during the estimation process on some stocks, the six GARCH family models(SGARCH-norm, SGARCH-std, SGARCH-sstd, EGARCH-norm, EGARCH-std, EGARCH-sstd), and LSTM, TCN, TCN (attention) based on deep learning framework are compared in terms of the results of the three-class classification tasks on the test set of each stock. As shown in Table 1, although the models based on TCN and TCN (attention) are slightly inferior to the GARCH models in terms of *Accuracy* metric, the performance in *Recall* metric of these two models on category A and category C are significantly higher than all the six GARCH family models. Although the LSTM model has the best *Recall* performance among all models in category A and category C, its overall performance in *Accuracy* on the three-classification task is much lower than all the other eight models.

Table 1. Part of confusion matrix of three-class classification.

| Model | Class A | Class B | | Class C | Accuracy |
|---|---|---|---|---|---|
| | Recall | Precision | Recall | Recall | |
| SGARCH-norm | 0.000157 | 0.738363 | 0.999785 | 0.000152 | 0.738253 |
| SGARCH-std | 0.000298 | 0.738412 | 0.999772 | 0.000211 | 0.738292 |
| SGARCH-sstd | 0.000235 | 0.738429 | 0.999491 | 0.000208 | 0.738132 |
| EGARCH-norm | 0.006455 | - | 0.963065 | 0.031123 | 0.717237 |
| EGARCH-std | 0.031414 | - | 0.947546 | 0.044524 | 0.708483 |
| EGARCH-sstd | 0.026612 | - | 0.964196 | 0.021047 | 0.718937 |
| LSTM | 0.236303 | - | 0.536982 | 0.333191 | 0.459041 |
| TCN | 0.135237 | 0.733206 | 0.83022 | 0.105678 | 0.640502 |
| TCN (attention) | 0.058021 | 0.738631 | 0.920458 | 0.068285 | 0.694751 |

For each of the nine models, the situation in which none of the sample of a certain stock is classified as category A sample or category C always exists. So, the averaged *Precision* can not be calculated for the nine models in terms of category A and category C. In order to continue to compare the nine models in the metric, this study counts the number of stocks on which the corresponding model achieves the highest prediction *Precision* in category A, B, and C. As shown in the following table, UHF price change models based on LSTM, TCN, and TCN (attention) structure are still the best three models. Taking the lower *Recall* of GARCH family model in Table 1 on category A and category C into consideration, this study found that the six GARCH family models are highly affected by the large proportion of sample with price change 0, and the six GARCH family models are more inclined to simply classify the samples as category B. Therefore, TCN and TCN (attention) models perform better than GARCH models on three-class classification tasks in general.

Table 2. Three-class classification precision results of each model.

| Model | Class A | Class B | Class C |
|---|---|---|---|
| SGARCH-norm | 9 | 2 | 13 |
| SGARCH-std | 3 | 3 | 8 |
| SGARCH-sstd | 3 | 3 | 2 |
| EGARCH-norm | 13 | 2 | 7 |
| EGARCH-std | 6 | 4 | 5 |
| EGARCH-sstd | 5 | 9 | 4 |
| LSTM | 22 | 38 | 16 |
| TCN | 17 | 32 | 33 |
| TCN (attention) | 25 | 13 | 23 |

Table 3. Part of confusion matrix of five-class classification.

| Model | Class 1 | Class 2 | | Class 3 | Class 4 | Accuracy |
|---|---|---|---|---|---|---|
| | *Recall* | *Precision* | *Recall* | *Recall* | *Recall* | |
| LSTM | 0.144859 | - | 0.536982 | 0.195811 | 0.166845 | 0.433289 |
| TCN | 0.060193 | 0.733206 | 0.83022 | 0.040967 | 0.092275 | 0.625623 |
| TCN (attention) | 0.029599 | 0.738631 | 0.920458 | 0.034279 | 0.039488 | 0.688268 |

As shown in Table 3, this study also separately compares the averaged confusion matrices of LSTM, TCN, and TCN (attention) in the five-class classification task. It can be found that compared with the TCN model and the TCN (attention) model, the LSTM model is more sensitive to price changes, but its overall performance is also much more unstable. The prediction *Accuracy* of LSTM model on the five-class classification task is also much lower than the TCN model and the TCN (attention) model.

## 4.Conclusion

This study focuses on establishment of the discrete dynamic distribution model for UHF price changes, and compares the traditional GARCH family models with three dynamic distribution models based on LSTM, TCN and TCN(attention). The empirical research on 99 constituent stocks of SZSE 100 Index prove the robustness of TCN structure. For future research, the problem of unbalanced sample will be taken into consideration and the different UHF models will be ensembled to provide a more stable prediction for different financial securities.

## Acknowledgements

This research was partly supported by National Natural Science Foundation of China (No.71771204) and the Fundamental Research Funds for the Central Universities.